# Entropy and Economics

Martin Pomares C. [1]
http://orud.org/0000-0003-4994-0573

## Introduction

Entropy is a very useful concept from physics that tries to explain how a system behaves from a point of view of the thermodynamics. However, there are two ways to explain entropy, and it depends on if we are studying a microsystem or a microsystem. From a macroscopically point of view, it is important to describe if the system is a reversible system or not. However, form the microscopically point of view, the concept of chaos is related to entropy. In such case, entropy measures the amount of disorder into the system. Otherwise, the idea of connecting at the same time the analysis of the macro and micro system with the use of entropy it is not very common.

In addition, the study of complex systems implies to consider several approaches to understand the behavior of such systems. In that way, the economy is an interesting universe of study were the application of entropy is becoming very new in order to apply thermodynamics to understand form the point of view of the physics how a macro or a micro system can be behave.

## Part A: Use of the entropy in the macro-economy

One way to apply the concept of entropy into the economy is when the macro-economy is studied from the perspective of the production function. In that case, the production function should be studied form the non-neoclassic approach. With that purpose, the entropy becomes a new tool of analysis.

In any kind of market, the basis of his existence is not only the flow of the money, but also how the flow of products and goods that are inter connected in the market. Such products has a span life no very long that has some influence in the price of themselves. In the products, prices and span life are variables determinants of the consumption into the markets (ScalesAvery, 2012). Those variables are also determinants in the quality of the markets, which also is related to the entropy of the system, and it is reflected in the production function Cobb Douglas (De Pascale, *2012*).

In 1989, J. Rifkin published a paper titled "Entropy into the Greenhouse World" where it was shown how the second law of thermodynamic impact on the economical process that affect the environment (De Pascale, *2012*).

The first and the second one law of the thermodynamics are respectively the law of conservation of energy, and the low of reversible process or "low of entropy". This last one let us determine the direction of thermodynamically processes based on the variations of the entropy that the system experiences. In consequence, the entropy measures the degree of irreversibility of the system (De Pascale, *2012*).

According to Raine, A. et al. (2006), economist argue that "the evolution of economic systems is a process of increasing structural complexity driven by the production of new knowledge. But much debate remains about how and why this might be. For a long time, the micro foundations of this approach have been sought by way of analogy with biological evolution". But also in "the theory of economic evolution remains an open question". Otherwise, the author pretend to consider entropy in order to measure distribution of wealth in a society from a macroscopic point of view but no fluctuations in financial markets as it can be also regarded in future studies as well.

---

[1] Researcher at Science and Engineering Faculty, UNAN-Managua. Nicaragua. Central America.

Considering the ideas before presented, the author has developed a study in a macro-economic system for Nicaragua, in Central America. Where the main objectives were: To estimate the production function or entropy for the macro-economic system of that country; and to formulate a production function that let us analyze the irreversibility evolutive of the system[2]; also analyze the evolution of the macro-economy of that country in between 2000 and 2010 regarding the entropy estimated as a production function proposed by Pomares, M. (2019).

**Theoretical framework**

The entropy ($S$) can be estimated as a function of the heat interchanged ($\delta Q$) in between the system and the surroundings when the system behaves to a constant temperature ($T$) defined by[3]

$$S = \oint \frac{\delta Q}{T} \tag{1}$$

From the Econophysics approach, the economical temperature $T$ is proportional to the GDP per capita for a macro-economic system (Chakrabarti et al., 2006). As a consequence, $1/T$, is the integrant factor which depends on the average of the capital ($E$), such as $E = \alpha NT$. Where N is the number of agents of the economic system that is studied, and y $\alpha$ is constant of proportionality (Chakrabarti et al., 2006). In other side, in order to consider the exact differential of entropy from the before definition in the equation (1), it is possible to substitute the differential of heat (interchanged heat) ) $\delta Q$ regarding the first law of thermodynamic, it is possible to rewrite the exact differential as

$$dS(T,V) = \frac{\partial S}{\partial T}dT + \frac{\partial S}{\partial V}dV = \frac{1}{T}\frac{\partial E}{\partial T}dT + \frac{1}{T}\left(\frac{\partial E}{\partial V} + p\right)dV \tag{2}$$

Where $p$ is the economic pressure that the system experiences. The concept of entropy was firstly introduced in economy by N. Georgescu-Roegen (1974), and recently by D. K. Foley y J. Mimkes (1999[4]). From the point of view of the Econophysics it is understood over the concept of **production function** or **economic entropy**, defined as

$$S(N_k) = \ln P(N_k) \tag{3}$$

---

[2] The analysis of the irreversibility of the system considers some ideas according to Perrings, Ch & Brock, W. (2009)
[3] According to Clausius, R. (1867).

Where $P(N_k)$ is the polinomial distribution[4] of $N_k$ goods or categories. In stochastically systems, the entropy substitute the production function of Cobb Douglas into the standard economy. In such way, there are several reasons to consider, such as: First, The entropy is a natural function without additional parameters like the elasticity that uses the Cobb Douglas function. Second, the entropy and the Cob Douglas function are adjusted very well to the same data that implies the same behavior for those both functions. Third, the entropy characterize the change of distribution of products and finances during the production process and commerce. In the commerce of products, it is generally finished when the equilibrium is reached in between products and finances.

When the variation of entropy in the system is positive, it is regarded a distribution of goods and services creating a disorder, in change, if the variation of entropy is negative, it will be understood that into the system products and money are being collected, creating order into the system.
However, it was regarded by the author to study the entropy for a macro-economic system from the point of view of macroscopic dependence or macro models as it is presented in the equation (2), regarding as approximation a closed economy.
Complex systems always experiences some dissipative structures as consequence of his own organization that depends on external restrictions and this restrictions depends on values such obligates to the system to reach states far of the equilibrium. Furthermore, the system itself evolves to a new complexity order interchanging matter, energy and information. Such evolution operates through the denominated "order by fluctuations" that corresponds to periodic fluctuations that the system experiments around his equilibrium. In intelligent systems and high complexity this phenomena let a constant "self-organization" (Gómez, 1977).
According to Goodwin (1976), those fluctuant processes are also denominated "Limit Cycles", that in the case to be applied to the economy are studied form the point of view of mathematical models of the "equilibrium theory" (Gómez, 1977). In addition to the before exposed, a characteristic of the equilibrium is the maximal entropy (Kåberger y Månsson, 2001). However, in the disequilibrium state, some processes conduct to a certain production of entropy. Where some systems can interchange entropy allowing that, the entropy stay constant in the system (Kåberger y Månsson, 2001).
In addition, the application of the concept of entropy and the laws of thermodynamics to the economy and the ecology states others perspectives of application braking through the neoclassical concepts of the economy.
According to GarretHardin (1993), a neo-Malthusian of the XX century, the second law of the thermodynamic is the fundamental basis that stablishes the limits to reach the level of population sustainable (Schwartzman, 2007). In equivalent way Georgescu-Roegen unify the concept of entropy to the Schumpeterian considerations, regarding with high emphasis the irreversible transformations more than the equilibrium points of the system and its regularities (Carpintero, 2005). Because of the human development, an entropic increment is generated because of the natural resources, generally of the irreversible character. In this case, the entropy measures it selves the limits of the sustainable development (Schwartzman, 2007). Otherwise, it is important to remark the existence of states of disequilibrium, which are more related to inflationary processes. The materialist vision of the inflation, which anywhere a monetary phenomenon, establishes that the inflation reflects the imbalances in the markets of products and works. This point of view is not consistent to regards the excess of demand by money as a source of pressure for the inflation (Bardsen et al., 2005).

---

[4] The *polinomial distribution* is defined as $P(N_k) = \frac{N_0!}{N_1! \cdots N_k!} \cdot q_1^{N_1} \cdot \ldots \cdot q_k^{N_k}$, y $N_k$ son las diversas categorías para el cual $N_0 = N_1 + \cdots + N_k$.

**Method of use of entropy as a production function**

For the macroeconomic analysis, it is possible to rewrite the exact differential of entropy in order to set up as a function of the capital ( E or GDP capital), the volume (V) which is the volume of the worker population for a country or a region. The economic pressure can be rewrite too as a function of the work and volume as: $p = W/V$. In addition, regarding that the energy is proportional to the temperature, and then the concept of work (*W*) will be understood as the stock of capital. Then after the integration, the entropy can be written as:

$$S(E,V) = \alpha N \ln E^2 + W \ln V = \ln[E^{2\alpha N} V^W] = \ln \xi \tag{4}$$

Where $\xi = E^{2\alpha N} V^W$, and $\alpha$ is a constant of proportionality and calibration that establishes a relation between the temperature and energy (Chakrabarti et al.,2006). The equation (4) represents a production function for macro economy for a system that depends of the capital and of the worker population active. In this case, as a first approximation $\alpha$ obtain the value of one regarding that the temperature and the energy have a monetary unity, and the unique agent studied is a macroeconomic system where the capital corresponds to the GDP.

In order to estimate the "*entropic elasticity*" or elasticity in the simple way, we have the following equation:

$$\frac{dS}{S} = \frac{d\xi}{\xi \ln \xi} = \frac{1}{\ln[E^{2\alpha N} V^W]} \left[ 2\alpha N \frac{dE}{E} + W \frac{dV}{V} \right] \tag{5}$$

The entropic elasticity measures the contribution of the population active worker with respect of the GDP; it is to say the propensity of production regarding the active population worker. The entropy works as a production function where technology is implicit, and for that reason only depends on the capital (GDP). It is because the development carried out by the active worker population depends on the technology to make the production itself. In addition, the entropic elasticity measures the relativity productivity of entropy in a macro economic system, which will be the sense of a non-classic of the use of the production function denominated *entropy*.

The author presented a paper titled "**Entropía en los Modelos Macroeconómicos: Otra perspectiva de Función de Producción**"[5] (<http://orud.org/0000-0003-4994-0573>), that was published at REICE: Revista Electrónica de Investigación en Ciencias Económicas.
. In this investigation, the concept of entropy as a production function is studied in between 2000 and 2010 for the GDP of Nicaragua, in Central America. The elasticity is compared with the entropy proposed in this section as a production function. The results obtained can support the hypothesis that entropy can be used as new kind of production function but of course, the data obtained in that research has the same behavior but not the same numerical values. It means that there are more researches needed into macro-economy in that topic that belongs to the Econophysics. Otherwise, the entropic analysis let us understand if the system is reversible or not. In addition, this works is another study in order to link the thermodynamics to the economy according to Georgescu-Roegen (1974).

**Part B: Use of the entropy in the micro-economy**

In micro-economy, one of the most important aspects to study is the behoviour of consumption, which is related to production, goods and services. Due to economy is a complex system, the modelling of that kind of systems is not so easy because it will depends on what kind of variables it is needed to study. For this reason is that cellular automata is a kind of modelling of complex system adequate to analyze how a system itself evolves. However, regarding also that consumption is an interchange when someone else is buying something; the game theory takes some importance in order to complete the analysis of the behavior of markets.

The theory of cellular automata takes its origins with Von Neumann (between 1940 and 1970) with the purpose of study automata or machines that can self reproduces, then with the with J. H. Conway (between 1970 and 1983) in order to study the game of life. After, since 1983 S. Wolfram developed a concept to study that he denominates "The statistical mechanics of cellular automata". From Wolfram, we know the mathematical concept of cellular automata with several applications to the science (Pérez, 2012).

Otherwise, the model of cellular automata can be understood as a kind of a graphical picture, where as a system, a cellular automata is composed by states, rules and surroundings or neighborhoods. These states can be plotted by cells in a diagram 2D, which could be binaries states, but also we can set up the colors black and white for such states or others. The white color can represent the binary state 1, and the black color can represent the binary state 0, or vice versa. The neighborhood corresponds to mathematical form that a state dynamical pass from one state to other. And the rules corresponds to the mathematical way in which an state pass dynamically from one state to other nearly or far away depending on how is defined the interaction between states (Fig.1).

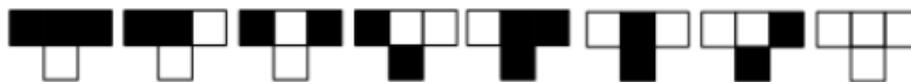

Fig.1 Evolution of the states of cellular automata according to the rule of evolution of the states of cellular automata.

---

[5] Available at: https://revistacienciaseconomicas.unan.edu.ni/index.php/REICE/article/view/390

Otherwise, the most remarkable use of cellular automata is the qualitative analysis by Wolfram (1984) and Li & Packard (1990), and the quantitative analysis by Weunsche (1998) using the concept of entropy of entrance, and Langton (1990) who use the exponents of Lyapunov. However, into the theory of cellular automata are also studied the emergent behavior like self-regulation, hierarchical and non-hierarchical structures. Moreover, more recently into the economy, cellular automata are used mostly to analyze the marketing in between systems. In similar way, the author regarding cellular automata connected to the concept of entropy studied the market for a micro-economic system as it was presented in the paper titled "**Cellular Automata in the Economy: Application to the Behavior Theory of the Consumers.**"[6]

In addition, some models of cellular automata are used to study of complex systems, which according to Meagher. K. & Rogers, M. (1997), in 1989, Santa Fe, a group pf physicist and economist were studding about the connection in between both disciplines. Therefore, the memory of that conference was titled *The Economy as an Evolving Complex System* which contains papers about the scope of technics that includes chaos theory, neural networks, and cellular automata.

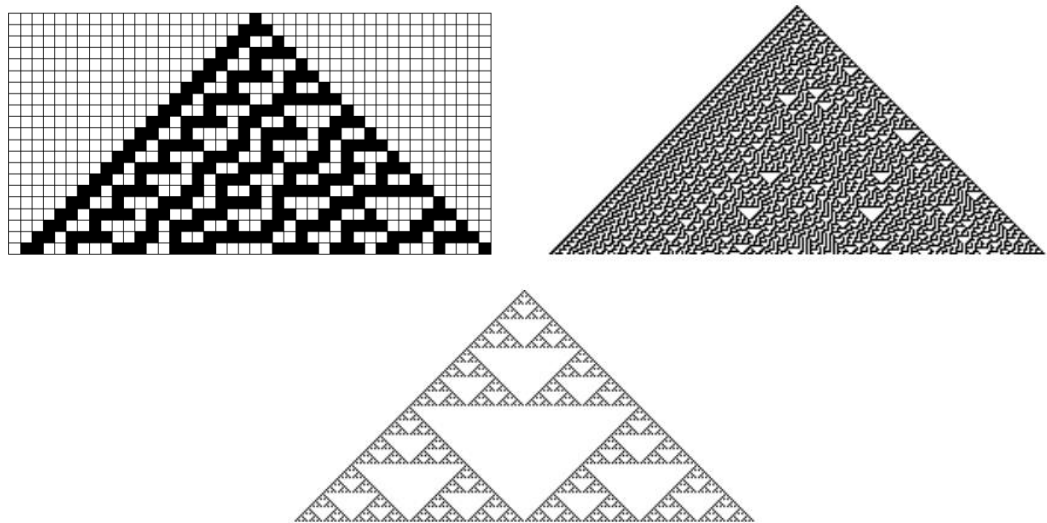

Fig. 2. Rule 30 with a cellular automata of 43 cells according to Wolfram (left up); rule 30 with a cellular automata of 257 cells (right up). Rule 90 (center below).

**Mathematical model for the analysis of a system using cellular automata**

There is not so much emphasis in the modelling of the theory of preferences revelated into the economic theory. In such case the present model considers to study the dynamics of the theory of consumption through a graphic of prices ($p_i$) versus goods ($x_i$) which in the vectorial space is represented a state which corresponds to the rent ($w_i$) (Ver Fig. 3). Such rent will be represented in a quantized way by a state of cellular automata in an array of 2D. The dynamics of the rent correspond to the dynamic of the cellular automata evolving in their states, and it will change of state in a configuration space (Fig. 3). The area of each state corresponds to the minimal rent in that a consumer select spend into a good. Regarding the fact that the selection will be optimal (of course it is an assumption *a priori*), when the dynamic of rent increase also increase the possibility to obtain de adequate preference or satisfaction. Therefore, and in a simultaneity to the model of cellular automat the model of a *Gibbs distribution with measurement of entropy* is simulated in order to use the concept of entropy to measure nonsatisfaction into the system as consequence of the dynamic of the cellular automata. Besides, to study the

---

dynamic of cellular automata it is considered into the segregation model of Schelling the modulation in between two agents or goods.

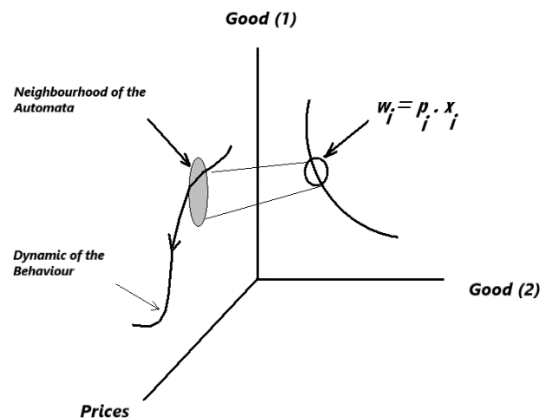

Fig. 3. Pictorial idea of the use of cellular automata for this research. The configuration space of Prices versus Goods will be useful to understand the dynamic of consumption (Pomares, 2022).

**Schelling´s Segregation Model**

In order to applicate the Schelling´s model, it is necessary to regard *N* agents that develop local interactions; it is to say with his nearly neighbors that let us to consider the area of local interactions as a neighborhood of ratio *r*. It is important to start from the assumption that the agents are represented by "D cells located in a rectangular reticula, where each one has a neighbor to watch, and each one has neighbor of the same type in his neighborhood. A restriction is that each agent has eight neighbors to watch. In order to use cellular automata, each agent represents a good ($B_i$), and for commodity of the study two different agents will be simulated. We regards a tolerance represented by a positive number $m \in \{1, ..., 2r + 1\}$, which determine the maximal number of neighbors non fancy that the agent is able to admit. The tolerance is a parameter that can be understood as a threshold of nonsatisfaction that each agent admits in his neighbor. For the application of cellular automata, if the agent poses les neighbors that the parameter THRESH (threshold), then those are no satisfied, and as a consequence interchange his positions in order to restart their observations (Benito & Hernández, 2007).

The individual utility, which measures in a binary way the level of satisfaction in individual way is generated by its neighborhood. The Set of information of each agent *i*, its neighborhood of ratio r in the state *t* denotated by $V^t(i,r)$ is equal to an element of $\{0,1\}^{2r+1}$ focused in *i* and the utility of the agent *i* in the state *t* is represented as bellow (Benito & Hernández, 2007):

$$U^t(i) = \begin{cases} 1 \; si \; |\{j \in V^t(i,r) \; tal \; que \; j \neq i \}| \leq m \\ 0 \; si \; |\{j \in V^t(i,r) \; tal \; que \; j \neq i \}| > m \end{cases}$$

In addition to the before explained, the utility function set that each member corresponds only the number of neighbors as much as prefer as are not preferred, that implies that to the agents are substantial more the composition than the configuration. It is a kind of a rule of interaction. Then, the dynamic is an iterative process where the agents selects the best answer given by the set of local information (Benito & Hernández, 2007).

Besides, of the Schelling´s model, a model form statistical mechanics is needed in order to study the entropy of the system as part of the behavior of consumption. Therefore, according to Dragalescu and Yakovenko, (2000), and Dragalescu, (2003), it is possible to develop a statistical analysis to a system with multiple agents composed by N players, each one of them with an amount of money $m_i$.

One of the characteristic of this model is that the **N** players are committed in a lottery where draw a pair *i; j* of them and then randomly is decided who of them win an amount **m**, and who of them loss the same amount. In addition, If the agent that loss cannot afford the payment **dm**, the play is canceled. But, it is necessary to estimate which will be the probability that we can find agents with the trend of amount of money **m** that results stationary against this kind of transactions. In order to obtain a good estimation, a probability is calculated with the Gibbs distribution used in the statistical mechanics described by $P(m) = Ae^{-\beta m}$, where $\beta = 1/T = N/M$ represents the temperature that for an economical system is the amount of money average by agent.

However, it is needed to know the entropy of the system, because the amount of disorder that the statistical process is obtained from the entropy that in this case, as our model uses discrete variables, the entropy can be defined by $S = \log(M)$.

In addition, to the theory presented before, the present model study the dynamic of the consumption theory. A graphic of prices (*p$_i$*) versus goods (*x$_i$*) which in the cartesian vectorial space it will be represented by a cellular automata 2D where each state or cell represents the rent (*w$_i$*) which area describe the minimal value of the rent connected to the good. Then, the Segregation Schelling´s model of cellular automata is used in order that fulfil form the mathematical point of view with the adequate characteristics to study the dynamics of states.

One assumption is that in our model there are two agents or goods. In the measurement of nonsatisfaction. The statistical estimation of nonsatisfaction caused by the dynamics itself is measured. According to the second law of the thermodynamics, there is a certain degree of entropy. In order to measure that entropy, it was used the model of **random interchange game considering a Gibbs distribution**, as a part of the selection process. Moreover, two players (N = 2) are considered.

In the Fig. 4 are represented some results obtained from simulation considering the model proposed before. This picture shows the evolution of estates considering the selection of states of satisfaction and nonsatisfaction for a good or other of the agents evaluated. Something similar to phase space in classical mechanics. The valid states or of preferred decision are presented in red color, regarding the initial state in the upper left corner for all the items of the Fig. 4. The blue color represents an empty state it is to say without value. All the red reticula represents the dynamic followed by the automata.

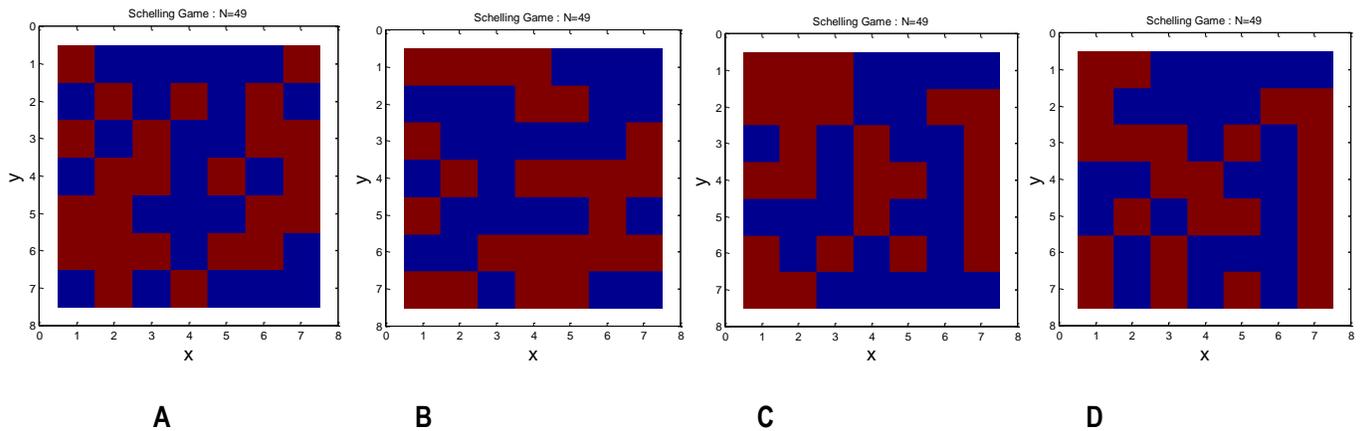

A      B      C      D

Fig. 4. Evolving of the states of the cellular automata for several *steps*: A =57, B = 70, C = 91, D = 135 (Pomares, 2022).

In the graphic of the Fig. 5 it is possible to observe the evolution of the entropy, it is to say in between satisfaction (entropy equals to zero) and nonsatisfaction (maximal entropy equals to 0.63 units). Furthermore, the pathways of the red cells indicates the places of the dynamics of the rent where there are optimal decisions of the consumer.

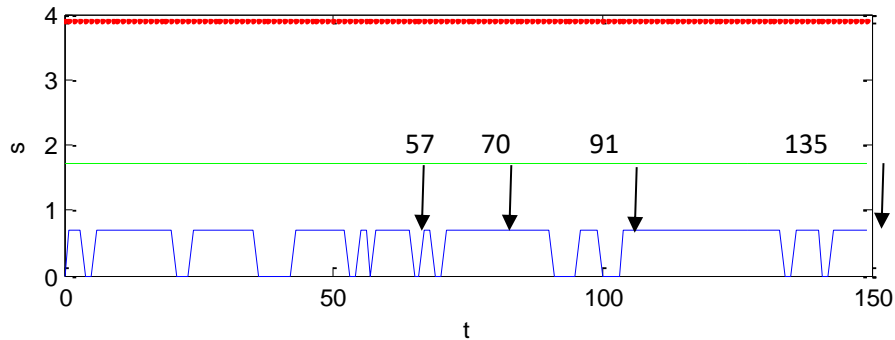

Fig. 5. Diagram of Entropy (S) versus time of the probabilistic estimation (Pomares, 2022).

**Concluding Remarks for this Part B**

The model presented in this part study the dynamic of preferences reveled in a new space of configurations of prices versus goods with the purpose of to obtain a graphical idea of the dynamics of how evolves the state of satisfaction and nonsatisfaction according to Fig. 4. In other case, it is regarded the measurement of entropy of the system over predefined considerations showing a maximal entropy stable. However, from the micro-economy approach, it is needed more studies in order to applicate thermodynamic and other background from physics to economic systems. Besides, it is important to regard, how interconnect the physics theory of macro-systems to micro-systems and vice versa.

**Bibliographical References**

**Part A**
Scales Avery, J. (2012) Entropy and Economics.Cadmus Journal. Volume 1, Issue 4.

Jaynesy, E. T.(1991) HOW SHOULD WE USE ENTROPY IN ECONOMICS?

Chakrabarti, B. K., Chakrabarti, A., Chatterjee, A., (2006) Econophysics and Sociophysics.Trends and Perspectives.Willey-VCH.

Bardsen, Gunnar., et al. (2005) The Econometric of Macroeconomics Modelling.

De Pascale, Angelina (2012) Role of Entropy in Sustainable Economic Growth.International Journal of Academic Research in Accounting, Finance and Management Sciences.Volume 2, SpecialIssue 1, pp. 293-301.

Raine, A. et al. (2006) The new entropy law and the economic process. Ecological Complexity. ELSEVIER. Volume 3, Issue 4, Pages 354-360. https://doi.org/10.1016/j.ecocom.2007.02.009

Perrings, Ch & Brock, W. (2009) Irreversibility in Economics. Annual Review of Resource Economics. Vol. 1:219-238.

Clausius, R. (1867) The Mechanical Theory of Heat: With Its Applications to the Steam-Engine and to the Physical Properties of Bodies; Hirst, T.A., Ed.; John van Voorst: London, UK; ISBN 9789353740962.

Goodwin, B. C. (1976) Estabilidad biológica. En "hacia una biología teórica". Editado por C. H. Waddington y otros. Alianza Editorial Madrid. Pág. 424.

Kåberger, T y Månsson, B. (2001) Entropy and economic process ⁓ physics perspectives. Ecological Economics, 36. 165 – 179.

Gómez G., L. J. (1977) La Entropía y sus relaciones con la Economía y la Ecología. Primer coloquio sobre termodinámica y energía. Facultad de Ciencias Humanas y Económicas. Universidad Nacional. Medellín.

Schwartzman, D. (2007) The Limits to Entropy: the Continuing Misuse of Thermodynamics in Environmental and Marxist theory. In Press, Science & Society.

Hardin, Garrett J. 1993. Livingwithinlimits : ecology, economics, and population taboos. New York: Oxford UniversityPress.

Carpintero Redondo O. (2005). El desafió de la bioeconomía en Revista Ecología Política Nº 30, Barcelona.

PWT 7.1. Disponible en https://pwt.sas.upenn.edu/php_site/pwt71/pwt71_form.php

Pomares, M. (2019). Entropía en los Modelos Macroeconómicos: Otra perspectiva de Función de Producción. REICE: Revista Electrónica de Investigación en Ciencias Económicas. Vol. 7. No.14.

**Part B**

Pérez López, Andrés (2012). Autómatas Celulares. Development of a SuerCollider3 Class.

Reyes Gómez, David Alejandro (2011) Descripción y Aplicaciones de los Autómatas Celulares. Departamento de Aplicación de Microcomputadoras. Universidad Autónoma de Puebla (UNAM). México.

Meagher. K. & Rogers, M. (1997) Networks, Spillovers and Models of Economic Growth.

Benito, J.M. And Hernandez, P. (2007) Modelling Segregation through Cellular Automata: A Theoretical Answer. Instituo Valenciano de Investigaciones Económicas, S.A.

Heymann, D., Perazzo, R. y Zimmermann, M.G. (2011) Modelos económicos de múltiples agentes. Una aproximación de la economía desde los sistemas complejos. WORK IN PROGRESS.

Dragalescu, A. and Yakovenko, V. (2000) Statistical mechanics of money, Eur. Phys. J. B 17, 723{729.

Dragalescu, A. A. (2003) Applications of physics to economics and finance: Money, income, wealth, and the stock market. e-print arXiv:cond-mat/0307341v2.

Pomares, M. (2022). Cellular Automata in the Economy: Application to the Behavior Theory of the Consumers. Science Publishing Group. Economics. Volume 11, Issue 1, March 2022, Pages: 43-48. Received: Jan. 20, 2022; Accepted: Feb. 16, 2022; Published: Feb. 25, 2022.